\documentclass[journal]{vgtc}                
\ifpdf
  \pdfoutput=1\relax                   
  \usepackage{graphicx}                
  \DeclareGraphicsExtensions{.pdf,.png,.jpg,.jpeg} 
\else
  \usepackage{graphicx}                
  \DeclareGraphicsExtensions{.eps}     
\fi%

\graphicspath{{figures/}{pictures/}{images/}{./}} 

\usepackage{microtype}                 
\PassOptionsToPackage{warn}{textcomp}  
\usepackage{textcomp}                  
\usepackage{mathptmx}                  
\usepackage{amsmath}
\usepackage{times}                     
\usepackage{cite}                      
\usepackage{tabu}                      
\usepackage{booktabs}                  
\usepackage{enumitem}
\usepackage{float}
\usepackage[utf8]{inputenc}

\onlineid{1607}

\vgtccategory{Research}
\vgtcpapertype{application/design study}

\title{Real-Time Visual Analysis of High-Volume Social Media Posts}

\author{Johannes Knittel, Steffen Koch, Tan Tang, Wei Chen, Yingcai Wu, Shixia Liu, Thomas Ertl}
\authorfooter{
\item
 J. Knittel, S. Koch, and T. Ertl are with University of Stuttgart. E-mail: firstname.lastname@vis.uni-stuttgart.de.
\item
 T. Tang, W. Chen, and Y. Wu are with State Key Lab of CAD\&CG, Zhejiang University. E-mail: \{tangtan,chenwei,ycwu\}@zju.edu.cn.
\item
 S. Liu is with Tsinghua University. E-mail: shixia@tsinghua.edu.cn.
}

\shortauthortitle{Knittel \MakeLowercase{\textit{et al.}}: Real-Time Visual Analysis of High-Volume Social Media Posts}

\abstract{Breaking news and first-hand reports often trend on social media platforms before traditional news outlets cover them.
The real-time analysis of posts on such platforms can reveal valuable and timely insights for journalists, politicians, business analysts, and first responders, but the high number and diversity of new posts pose a challenge.
In this work, we present an interactive system that enables the visual analysis of streaming social media data on a large scale in real-time.
We propose an efficient and explainable dynamic clustering algorithm that powers a continuously updated visualization of the current thematic landscape as well as detailed visual summaries of specific topics of interest.
Our parallel clustering strategy provides an adaptive stream with a digestible but diverse selection of recent posts related to relevant topics.
We also integrate familiar visual metaphors that are highly interlinked for enabling both explorative and more focused monitoring tasks.
Analysts can gradually increase the resolution to dive deeper into particular topics.
In contrast to previous work, our system also works with non-geolocated posts and avoids extensive preprocessing such as detecting events.
We evaluated our dynamic clustering algorithm and discuss several use cases that show the utility of our system.%
} 


\CCScatlist{ 
 \CCScat{K.6.1}{Management of Computing and Information Systems}%
{Project and People Management}{Life Cycle};
 \CCScat{K.7.m}{The Computing Profession}{Miscellaneous}{Ethics}
}

\teaser{
  \centering
  \includegraphics[width=\linewidth]{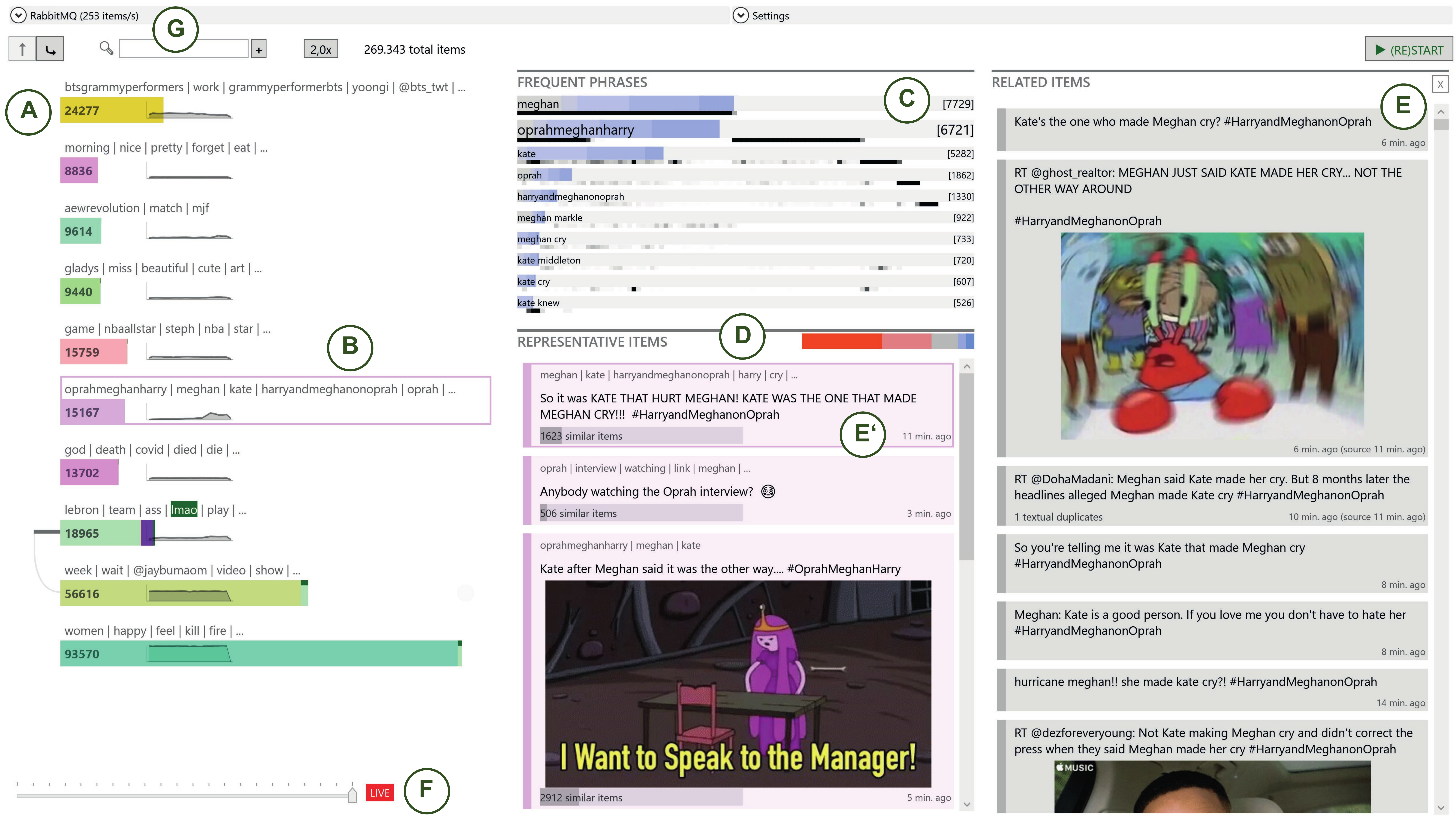}
  \caption{Overview of our proposed system applied to a real-time stream of tweets. \textbf{A:} Topical overview of the 270k posts in the sliding window. \textbf{B:} Selected topic of interest (ToI). \textbf{C:} Visualization of frequent phrases in the ToI. \textbf{D:} Stream of representative posts in the ToI. \textbf{E:} List of similar posts to the representative post E'.
  \textbf{F:} History slider. \textbf{G:} Dive into topics based on search query \emph{or} selected topics.}
  \label{fig:teaser}
}

\vgtcinsertpkg

\begin{document}


\firstsection{Introduction}

\maketitle


With the growing influence of social media platforms such as Twitter on society, the number of content creators, as well as the amount and topical diversity of published content on these platforms, has vastly increased.
People post about their daily experiences and opinions, businesses about their new products, and researchers about their latest findings.
Apart from everyday content, social media platforms are also a valuable source for breaking developments and news~\cite{Hu2012}, disaster management~\cite{Beigi2016,Thom2015}, and trading strategies~\cite{Oliveira2017}. 
Thus, several visual analytic approaches have been developed to facilitate the needs of various domain experts, including journalists, traders, and first responders~\cite{Wu2016,Chen2017}.

The sheer volume and speed of published posts pose a significant challenge.
Many approaches only support offline analyses, offer limited analytical capabilities, or cannot handle high-volume streams.
Only a few exist that support an online visual analysis of high-volume streaming data from social media.
However, they either rely on additional meta-data (e.g., voluntarily shared geolocation) or extensive preprocessing (e.g., event detection).

We propose a novel approach that aims to enable a visual analysis of high-volume social media streams in real-time, without constraints on additional meta-data or extensive preprocessing.
At its core, our efficient and explainable dynamic clustering algorithm groups incoming posts continuously while minimizing the amount of changes that each update would incur.
We run two clustering processes with different levels of granularity in parallel.
At the coarse level, we visualize the thematic landscape of the received posts with metaphors that are easy to comprehend and highlight what has changed after each update.
Analysts can select one or more topics to retrieve more information.
For such selected topics, we continuously extract and visualize frequent important phrases and their relationship to each other.
In addition, the fine-grained clustering process provides a digestible but diverse stream of recent posts related to this selection.
Analysts can dive deeper into topics either by specifying a search query or selecting relevant clusters to start a new session that filters the stream accordingly.

Our goal is to enable adaptive visual analyses irrespectively of the volume and velocity of the stream.
If analysts reside on the higher levels, they get a broad but still manageable overview of the data, and they can gradually increase the resolution to reveal more details, while still preserving their mental map.

In short, our contributions are as follow:

\begin{itemize}
\item We propose a dynamic clustering algorithm to enable the efficient clustering of fast-paced incoming streaming data. Our approach automatically determines the number of suitable clusters and minimizes the changes that each update incurs.
\item Based on our dynamic clustering algorithm, we developed a system for visually analyzing streaming social media data in real-time that scales to millions of posts even on a budget PC. It provides a highly interlinked visualization of current topics, allows analysts to dive deeper into specific topics of interest, and reduces the amount of visual changes to preserve the mental map.
\item We present a two-way parallel clustering approach to extract a filtered stream of representative posts. Our visualization of frequent phrases and the continuous extraction of such representative posts offer a comprehensive but still digestible summary of recent developments related to topics of interest.

\end{itemize}

\section{Related Work}

Most related to our work are visual analytic systems that support the analysis of streaming documents in real-time, which we discuss in Section~\ref{sec:onlineAnalysis}.
In Section~\ref{sec:topicModeling}, we compare our dynamic clustering algorithm with previous approaches.
In the subsequent three sections, we also briefly discuss less related \emph{offline} visual document analysis approaches to embed our work in a broader context.

\subsection{Topic Modeling and Clustering}
\label{sec:topicModeling}

Grouping documents into clusters or assigning topics to documents are popular ways to aggregate and visualize document collections.
We generally use the terms \emph{topics} and \emph{clusters} (of documents) interchangeably.
Non-Negative Matrix Factorization (NMF)~\cite{Lee1999} and Latent Dirichlet Allocation (LDA)~\cite{Blei2003} are frequently used topic modeling algorithms.
Spherical k-Means~\cite{Dhillon2001} is based on the popular k-Means clustering algorithm but replaces the Euclidean with the cosine distance, improving results on textual data.
Spherical k-Means++~\cite{Endo2015} adapts the k-Means++ strategy for initializing the centroids using the cosine distance.

While several techniques have been developed that incorporate temporal aspects~\cite{Mei2005,Peng2018,Wang2006,Zhang2010,Gao2011}, only a few support the online clustering of streaming data.
EvoBRT~\cite{Liu2015a} is an evolutionary multi-branch tree clustering algorithm based on Bayesian Rose Trees~\cite{Blundell2010} and Bayesian Hierarchical Clustering~\cite{Heller2005}, but is not efficient enough to handle large data sets. 
Several very efficient online versions of k-Means have been proposed that approximate the k-Means objective with different strategies: process each element only once and update the cluster centroids greedily after each element~\cite{Chakrabarti2006}, perform an approximated version of k-Means locally on batches and use these centroids as input for the global clustering~\cite{Ailon2009}, or perform clustering only on a cleverly chosen sample~\cite{Ackermann2010,Braverman2011}.
These online versions are fast, but in contrast to our dynamic clustering algorithm, they approximate the k-Means objective and the number of clusters is fixed.
Furthermore, only the first strategy of updating the centroids greedily leads to coherent clusters over time.
However, we integrated an optional sampling strategy to enable an efficient clustering that is only bound by the available memory.

\subsection{Topic-Based Document Analysis}

Early on, researchers facilitated clustering algorithms to scale the visual analysis of \emph{static} text documents, that is, without supporting streaming data or incremental updates.
Pirolli et al.~\cite{Pirolli1996} introduced the Scatter/Gather interface to explore large document collections with descriptive keywords and sample documents.
Later, several approaches~\cite{Dredze2008,Dou2011,Alexander2015} adopted LDA~\cite{Blei2003} to extract and visualize topics.
HierarchicalTopics~\cite{Dou2013} uses EvoBRT to cluster documents hierarchically.
TopicPanorama~\cite{Wang2016} lets analysts compare topics between different corpora with an interactive node-link diagram and is also based on EvoBRT.
El-Assady et al.~\cite{El-Assady2016} introduced topic-space views for visually analyzing conversations.

Carpineto et al.~\cite{Carpineto2009} emphasize the role of suitable cluster labels.
Chuang et al.~\cite{Chuang2012b} point out that it can be challenging to interpret automatically derived topics.
Alexander and Gleicher~\cite{Alexander2016} found out that while the quality of the topics seems to influence how easy it is for users to make sense of them, the visual representation has less of an effect.
As an alternative to completely automatic techniques, several works have investigated the use of interactive topic modeling for the analysis of document collections~\cite{Choo2013,Hoque2019,Park2018,Yang2020}.

Instead of using distinct topics, some approaches proposed visual spatializations to convey topical relationships or semantic similarity, either by projecting documents \cite{Heimerl2017,Kim2017,Wise1995} or keywords \cite{Endert2013,Paul2018,Knittel2020} onto two-dimensional maps. 

\subsection{Topic-Based Analysis of Time-Dependent Text Data}
The publishing date of a document represents an important metadatum that can help to identify and shape topics.
Some approaches process the complete data set once to extract topics solely based on the content and integrate the temporal metadata afterward in the visualization (\cite{Havre2002,Heimerl2016,Dork2008,Krstajic2011,Liu2009,Wang2012}). 
Others either utilize adapted clustering techniques that incorporate additional metadata such as the date into the clustering process itself (\cite{Cui2011,Cui2014}), or process the data set in bins, with each bin spanning a certain time range (e.g., one day), and then try to connect the resulting clusters between adjacent time steps afterward (\cite{Gad2015,Krstajic2013}).
However, all variants rely to a certain extent on a global view of the data set and do not easily support the online analysis of streaming data.

Several visual representations have been proposed to convey the temporal evolution of topics.
ThemeRiver~\cite{Havre2002} inspired many approaches to visualize the occurrence of topics over time in a streamgraph, resembling a river-like metaphor~\cite{Cui2011,Cui2014,Heimerl2016,Liu2009,Sun2014,Wang2012}.
CloudLines~\cite{Krstajic2011} visualizes the frequency of entities or events over time in rows.
Each column in StoryTracker~\cite{Krstajic2013} depicts clusters of news reports from the respective day, and visual connections between cells of neighboring columns reveal relationships between them.
Similarly, columns composed of keywords in ThemeDelta~\cite{Gad2015} represent specific date ranges and the brushing helps to trace the keywords over time.

\subsection{Offline Analysis of Social Media Data}
Shortly after the rise of new microblogging platforms such as Twitter, new approaches were developed to analyze static sets of social media data.
Vox Civitas~\cite{Diakopoulos2010} relates social media posts to the respective video that was commented on and visualizes extracted keywords over time.
I-SI~\cite{Wang2012} is an architecture that extends ParallelTopics~\cite{Dou2011} for analyzing social media data and latent topics using a high-performance computing cluster.
ThemeCrowds~\cite{Archambault2011} generates several tiles of multi-level tag clouds for each time span (e.g., days) to summarize twitter comments.
LeadLine~\cite{Dou2012} visualizes extracted topics in rows and integrates event detection and named entity recognition for visually analyzing text data.
SentenTree~\cite{Hu2017} was developed to summarize social media content while preserving the word order in a node-link graph.
Other approaches~\cite{Viegas2013,Wu2014,Chen2020a} focus on visualizing network aspects of posts such as the information flow.

Harvesting shared geolocations of posts is a popular way to visually aggregate data. 
TwitInfo~\cite{Marcus2011} lets analysts retrieve relevant tweets related to specified keywords and visualizes geolocalized tweets on a map.
SensePlace2~\cite{MacEachren2011} visualizes tweet volumes with a geo-heatmap for situational awareness.
TopoText~\cite{Zhang2018} aggregates and visualizes spatial topics on a map across multiple scales.
For a more thorough analysis of social media visual analytic approaches, see Wu et al.~\cite{Wu2016} and Chen et al.~\cite{Chen2017}.

\subsection{Online Analysis of Streaming Documents or Posts}
\label{sec:onlineAnalysis}

Processing and visualizing streaming data is challenging in several ways.
Data has to be processed with fast algorithms that support incremental updates~\cite{Rohrdantz2011RealT-26393}, and dynamic visualizations have to be developed that preserve the mental map of users~\cite{Krstajic2013a}.

D\"{o}rk et al.~\cite{Dork2010} introduced one of the first visual analytic systems to follow tweets of an ongoing event, which includes a ThemeRiver-inspired visualization conveying the temporal evolution of important topics.
Each stemmed word represents a topic, which limits the expressiveness of the topics, though.
Twitcident~\cite{Abel2012} automatically fetches relevant tweets for incidents that have been broadcast and provides a faceted search with enriched metadata, including named entity recognition.
Liu et al.~\cite{Liu2016} proposed a tree- and sedimentation-based visualization of topics in text streams that uses EvoBRT~\cite{Liu2015a} for clustering.
STREAMIT~\cite{Alsakran2011} and TwitterScope~\cite{Gansner2013} project items to a dynamic 2D plot.
STREAMIT applies a physical model to ensure the continual evolvement while new documents are received, supporting hundreds of documents.
TwitterScope~\cite{Gansner2013} projects tweets related to a keyword onto a map with MDS, either using cosine- or LDA-based similarity, and aims to maintain the relative position of nodes on each update that happens every minute.
Representing posts as dots has the advantage that all changes are visually apparent, but it does not scale well to hundreds of new posts each second due to the increased visual clutter.

Whisper~\cite{Cao2012} visualizes the diffusion of information on Twitter regarding different topics in real-time, with updates every five minutes.
The sunflower-like visualization in which tweets are represented as dots on a map integrates the geolocation of tweets.
The comprehensive ScatterBlogs system~\cite{Thom2012,Chae2012,Bosch2013} visualizes term usage anomalies from geolocated tweets and was later extended with an event detection algorithm, filter methods, and means to create and train classifiers interactively.
Their case study~\cite{Thom2015} shows that situation awareness domain experts consider the real-time analysis of social media content to be useful, e.g., for disaster assistance.
However, the percentage of geolocated tweets has steadily decreased in recent years, which renders approaches that rely on geo-annotations less useful.

Most similar to our work is StreamExplorer~\cite{Wu2018} which made it possible to visually analyze non-geolocated social streams with tens of thousands of posts on a budget PC.
In contrast to our work, it first detects important time periods (\textit{events}), and tweets belonging to an event can then be clustered based on GPU-assisted self-organizing maps (SOMs).
The weight vectors of the maps are initialized with the corresponding result from the previous run to create stable maps across updates.
Analysts can apply several interactive lenses, e.g., the word cloud lens, to investigate areas of the map and refine the SOMs interactively.
For building the tweet vector, each word is mapped to an index with a hash function to avoid a global dictionary, and the resulting vector is then projected to a lower-dimensional embedding with Random Sampling for efficiency.
Our pipeline exploits the sparsity of high-dimensional Bag-of-Words vectors and thus avoids a DR-induced loss of information.
In addition, our visualization of frequent phrases offers aggregations that are richer in context, the stream of representative posts ensures a comprehensive selection of relevant tweets, and analysts can increase the resolution of certain topics.

\section{Task and Design Requirements}
\label{sec:designGoals}

For many analysts and journalists, it is important to know what is currently happening on social media, what themes people currently talk about.
This need to stay informed about major new developments is also referred to as \emph{situational awareness}.
Apart from this more explorative task, the interest in \emph{monitoring} specific themes often increases if a major story is breaking.
In such situations, it can become challenging to quickly gain an overview of what has been posted and to extract new information, despite focusing on a single theme.

We, therefore, aim to tackle two main goals with our approach.
We want to support both the \emph{situational awareness} on social media and the specific just-in-time \emph{monitoring} of currently developing themes.
More specifically, we want to enable the following analytical tasks:

\begin{description}[align=left]
\item[\textbf{(T1) Overview:}]
Gain a continuous overview of major themes people currently talk about on social media
\item[\textbf{(T2) Details:}]
Learn more about specific interesting themes
\item[\textbf{(T3) Monitoring:}]
Constantly monitor specific themes to keep track of new developments
\item[\textbf{(T4) Dive-in:}]
Make specific themes the center of the analysis and increase resolution
\end{description}

We approximate \emph{themes} with automatically derived \emph{clusters} based on the textual content of each post, which has three important benefits.
First, the resulting clusters from topic modeling or clustering algorithms structure the content reasonably well to provide an overview and help with navigating the thematic landscape, even if they may not perfectly match the themes the analyst had in mind.
Second, structuring the data with content-based clustering imposes little restrictions with regard to the data that we can process (e.g., posts do not need to be geolocated).
Third, we avoid introducing additional uncertainties or delays caused by additional preprocessing such as event detection.
 
An important aspect of our approach is that we want to support the \emph{real-time} analysis of streaming data.
As a result, we need to deal with additional challenges compared to the analysis of static data sets.
We summarized these challenges into the following requirements that our approach should meet:

\begin{description}[align=left]
\item[\textbf{(R1) Efficiency:}] We rely on efficient methods that support interactive approaches on streaming data.
\item[\textbf{(R2) Flexibility:}] Our methods need to quickly adapt to incoming data because we can make only little a priori assumptions about the data that we are going to process. For instance, new important terms (e.g., hashtags) may appear that we would need to consider.
\item[\textbf{(R3) Consistency:}] The internal state should not change too much on updates to preserve the analyst's mental map and avoid confusion.
\item[\textbf{(R4) Sparsity:}] The extent and frequency of visual changes should be minimized to reduce the cognitive load.
\item[\textbf{(R5) Transparency:}] We need to communicate not only the state but also its changes, so that users can follow what is going on.
\end{description}

\section{Architecture}
\label{sec:architecture}

\begin{figure*}
  \centering
  \includegraphics[width=\linewidth]{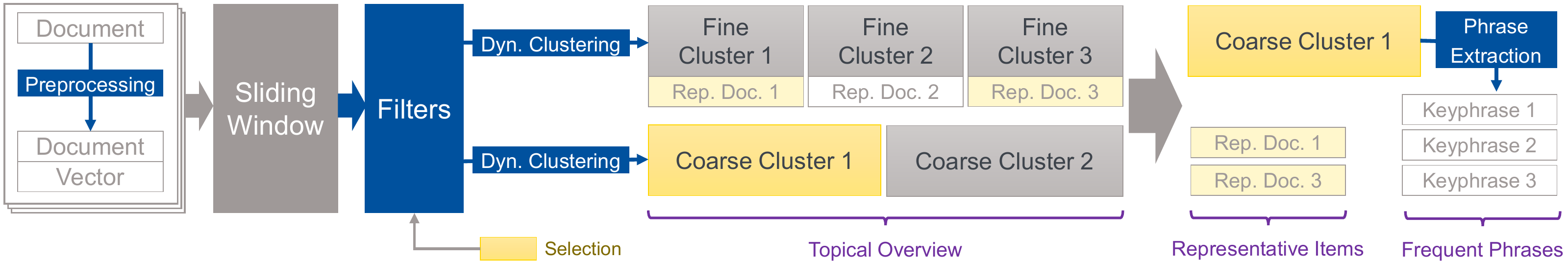}
  \caption{Architecture of our approach.
  Our system continuously collects social media posts and stores them in a sliding window.
  It runs two individual dynamic clustering processes in parallel.
  For each fine cluster, we find the representative item and match it to its closest coarse cluster (topic).
  Analysts can select one or more coarse topics.
  Frequent phrases in this selection will be extracted at regular intervals, and the associated representative items provide a diverse but manageable stream of relevant items.
  The cluster selection can act as a filter for a new session.
  }
  \label{fig:architecture}
\end{figure*}

Based on the requirements we set out in Section~\ref{sec:designGoals}, we developed a visual analytics approach that is powered by an \emph{efficient} and \emph{flexible} dynamic clustering algorithm to provide an overview of the currently posted content (T1), and to enable a detailed analysis of specific themes in a hierarchical manner (T2, T3, T4).
The system was programmed in C\#, runs under .NET 5, and its sole external dependency is ELSKE~\cite{Knittel21Elske} for extracting relevant keyphrases.

\subsection{Pipeline}

Figure~\ref{fig:architecture} depicts the architecture of our approach.
Our system continuously receives published posts and stores them and their derived vector embeddings in a sliding window with configurable size.
Each post is composed of a textual body, an optional language flag, and its publishing date.
Section~\ref{sec:preprocessing} details our preprocessing steps.
We apply our dynamic clustering algorithm (Section~\ref{sec:clustering}) to all items in this window at regular intervals of about one minute.

We establish two parallel and independent clustering processes with different levels of granularity (i.e., different thresholds for the maximum number of clusters).
By default, the first, \emph{coarse-grained} clustering does not extract more than $10$ clusters to provide analysts with an interactive \textit{topical overview} (T1).
The second process powers the diverse stream of representative posts with not more than $100$ clusters per default.
We set an upper limit of 10 for the number of main clusters so that we do not exceed the usual capacity of the analyst's short term memory, but both thresholds are adjustable.

We also call the coarse-grained clusters \emph{topics} and the more fine-grained ones \emph{subtopics}.
It should be noted, however, that topics and subtopics do not form a classical hierarchy since both clustering processes are independent from each other.
For each subtopic, we find its \emph{representative item}, that is, the post closest to the respective centroid.
Each post, therefore, has two cluster associations, one fine- and one coarse-grained, so each extracted representative item is also associated with exactly one topic.
Analysts can select one or more topics to retrieve additional \textit{details} (T2), including a stream of representative posts that are associated with the selection and extracted relevant keyphrases.
Such a selection of topics can be added as a new filter, which will create a new session layer that operates on the filtered stream.
Hence, with our layered approach analysts can interactively increase the resolution and adapt the specificity of their analysis (T4).

On every update, the frequent phrases will be updated and new representative posts may be added.
If a new subtopic appears or the representative item of a subtopic changes and is sufficiently different, the post will be added to the stream of representative items (T3).
The number of new items per update is limited because it correlates with the total number of subtopics.
These items offer a \textit{diverse} view of what is currently being posted since they originated from different clusters.

Compared to hierarchical clustering, our parallel clustering strategy ensures that both clusterings have reached their (local) minimum during the optimization; uncertainties do not accumulate across layers.
Furthermore, it is more straightforward to visualize and comprehend the dynamic changes of two individual, flat clusterings compared to a more complex dynamic hierarchy.

\subsection{Preprocessing}
\label{sec:preprocessing}

For each incoming tweet in the desired language, we create a sparse Bag-of-Words vector representation (BoW) as input for the clustering and for determining similar tweets.
We first remove URLs in the text, strip the \texttt{\#} from hashtags and remove the initial retweet markup if present (\textit{`RT @Username:'}).
We preserve username mentions because they often constitute helpful context.
Then, we tokenize the cleaned content (in lowercase) and assign each token its corresponding vocabulary index (we may need to add novel tokens to the vocabulary during this step).
We ignore stop words and punctuation characters.
For the final sparse vector, we set the value of the present token indices to their corresponding TF-IDF weight, and divide the vector by its length to retrieve unit vectors. 
We dismiss tweets that only contain stop words to avoid zero vectors.
For calculating the inverse document frequency, we use a random sample of tweets collected over several months.

More advanced neural network-based document embeddings may capture the semantics of each tweet better, but we opted for the TF-IDF~\cite{Salton1988} weighted BoW representation because it has several benefits that are important in our streaming setting.
First, embeddings usually operate on a fixed vocabulary, but we need to consider new terms and hashtags that often appear if a major story is breaking (\textit{flexibility}).
We would lose the ability to cluster the content based on such context-rich tags in case of a predefined vocabulary, which would degrade the quality and utility of our clustering.
Second, we want to visualize the clusters, so we strive for visually interpretable methods (\textit{transparency}).
With BoW vectors as input, the resulting cluster centroids can be interpreted as a weighted term list.
Hence, we can easily extract the terms with the highest weight to visualize the characteristics of each cluster.
Third, the BoW approach is very \textit{efficient}.
The inference on powerful language models with millions of parameters for creating the embeddings needs considerable processing time.
In comparison, our preprocessing pipeline can process more than 30,000 posts per second on a single core.
Furthermore, computing the dot product is much faster if one of the vectors is very sparse (see Section~\ref{sec:clustering}).
BoW vectors of tweets contain on average $20$ non-zero entries, whereas dense embeddings usually have hundreds or thousands of dimensions.
This has a measurable impact because calculating the distance is the defining hot path in the clustering algorithm.

\section{Efficient Dynamic Clustering}
\label{sec:clustering}

\begin{figure}
  \centering
  \includegraphics[width=0.6\linewidth]{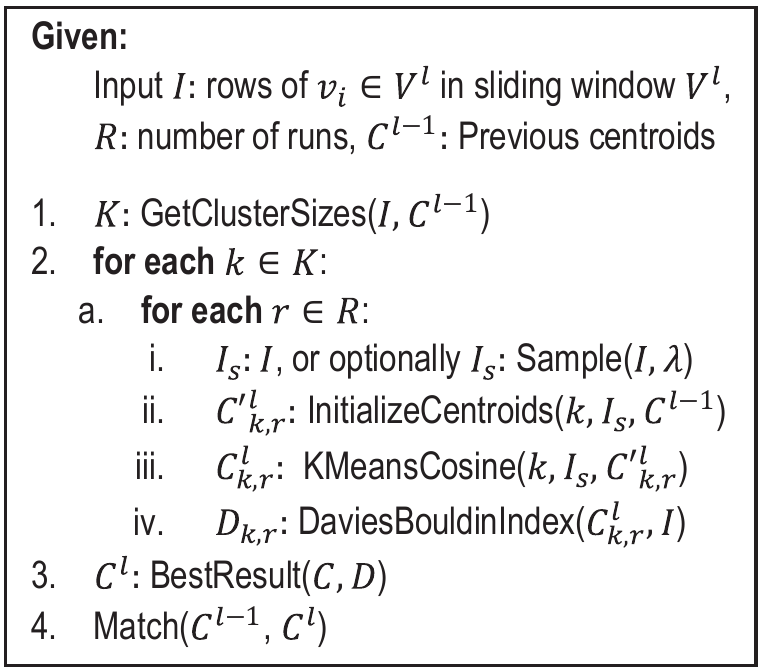}
  \caption{Main steps of our dynamic spherical k-Means++ algorithm.}
  \label{fig:algorithm}
\end{figure}

In this section, we describe our efficient dynamic clustering algorithm that we implemented in C\#. 
\emph{Dynamic} entails two important properties for the analysis of streaming data.
First, the algorithm should support incremental updates.
We want to reduce the amount of visual changes on each update (\emph{sparsity}) and we want to retrieve \emph{coherent} clusters over time.
Second, the algorithm should choose a suitable number of clusters within the provided constraints (i.e., the maximum number of clusters).
\emph{Efficient} means that our algorithm needs to be fast enough as it will be invoked at regular intervals.
We based our clustering algorithm on \emph{spherical k-Means++}~\cite{Endo2015} (sKMeans++) which performs competitively on document collections~\cite{lelu:hal-03053176}.
K-Means belongs to one of the fastest clustering algorithms, which motivates its use in the streaming setting.

The high-level idea of our dynamic version of sKMeans++ is as follows.
We take the centroids of the previous run into account when we set the initial centroids so that we get more coherent clusters over time.
We then run the optimization with different values of $k$ and choose the best result according to an internal evaluation criterion: the distance of each element to its corresponding centroid should be small, and the centroids should be sufficiently distinct from each other.
Figure~\ref{fig:algorithm} outlines the main steps of our algorithm.

The input $I$ comprises the document vectors of all items in the sliding window.
Compared to the previous run, some items may not be part of the set anymore, some will be new, and some may remain unchanged.
Let $k_{\text{min}}$ and $k_{\text{max}}$ be the desired minimum and maximum number of clusters, respectively.
We now have to determine which values of $k$ we would like to test and run the clustering with (\textit{GetClusterSizes}).
If it is the very first run, we set $K = \{k_{\text{min}}, k_{\text{min}} + 1, k_{\text{min}} +3, ..., k_{\text{max}}\}$.
Otherwise, $K = \{k_p, k_p + 1, k_p + 3, k_p + 6, ..., k_{\text{max}}\}$ where $k_p$ is the number of clusters from the previous run.
We increment the step size after every step to achieve a sublinear scaling regarding $k_{\text{max}}$.

The way we initialize the centroids $c_1, ..., c_k$ also depends on the previous run (\textit{InitializeCentroids}).
On the first run, we apply the sKMeans++ initialization strategy~\cite{Endo2015}:
we pick one element randomly as the first centroid $c_1$, and for each remaining $c_i$ we draw an element probabilistically based on its cosine distance to the nearest neighbor in the set of already chosen centroids $c_1, ..., c_{i-1}$.
If all distances are zero, the current set of centroids already cover all data items, so we stop the loop early and decrease $k$ accordingly.
Hence, the initialization will never return a clustering with duplicate clusters, even if this means that $k < k_{\text{min}}$.
On an incremental run, we first apply the old clustering to the new inputs and determine the set of $p$ non-empty clusters.
We then set $c_1, ..., c_{p}$ as the first $p$ initial centroids and determine the remaining centroids $c_{p+1}, ..., c_{k}$ with the initialization strategy outlined above.

Given these initial centroids, we perform the optimization loop until convergence (\textit{KMeansCosine}).
This is analogous to the Euclidean-based k-Means, with two exceptions.
First, for calculating the centroid vectors in the update step, we take the sum of all associated vectors and divide the vector by its length instead of calculating the arithmetic mean of the vectors.
Second, after the step of assigning items to their closest centroid, it can occasionally happen that we get empty clusters due to the fact that the cosine distance is not a metric.
If this happens, we just remove the respective cluster and decrement $k$ accordingly.

For each $k \in K$, we run the optimization process $R$-times to mitigate the impact of a bad initialization (the default value of $R$ is $2$).
We calculate the Davies-Bouldin-Index (DBI)~\cite{Davies1979} for each clustering result and return the clustering with the lowest score (\textit{BestResult}).
The DBI is an internal criterion for measuring the quality of a clustering.
We match a previous cluster $c^{l-1}_i$ to a current cluster $c^{l}_j$ if the majority of items associated with $c^{l-1}_i$ that are still in $I$ are now associated with $c^{l}_j$ and there is no larger group of previous items from a different cluster for which this also holds.

In the text domain, we have very high-dimensional vectors since the vocabulary can easily grow to hundreds of thousands of words, but the input vectors are usually sparse.
We, therefore, store all vectors and perform all calculations in a sparse format.
For unit vectors, calculating the cosine similarity is equivalent to taking the dot product, which we can exploit to speed up the computation because the number of entries that are non-zero in both vectors is often very small.

sKMeans++ usually converges fast, but the \emph{optional} sampling strategy further increases the efficiency so that we can cluster millions of documents within seconds.
One advantage is that we can apply any clustering to new, unseen data.
Thus, we can perform the clustering on a smaller subset and extrapolate the results to the complete data set.
Given a sample ratio $\lambda$, we pick $\lambda |I|$ rows randomly as input for the actual clustering run (\textit{Sample}).
However, we always use the complete data set when calculating the DBI.
Throughout this paper, we set $\lambda$ dynamically such that we run the optimization with at most 100,000 items to ensure fast response times also for millions of posts.

\section{Visualization Techniques}
\label{sec:applicationDesign}

As outlined in Section~\ref{sec:architecture}, our visual analytics system continuously receives hundreds of social media posts each second that we process with our parallel dynamic clustering strategy.
In this work, we focus on tweets, but the approach would also work with textual posts from other social media platforms.
Figure~\ref{fig:teaser} shows the user interface of our system.
On the left side, the \emph{Topical Overview} (A) visualizes the extracted topics, which we describe in Section~\ref{sec:topicalOverview}.
Analysts can select one or several topics of interest for additional details (B).
This will activate the \emph{Frequent Phrases View} (C) that contains a visual summary of the most important keyphrases in the selection, and the \emph{Representative Items View} (D) with a stream of diverse and relevant tweets.
We discuss both views in Sections~\ref{sec:frequentPhrases} and~\ref{sec:representativePosts}, respectively.

\subsection{Topical Overview}
\label{sec:topicalOverview}

\begin{figure}
  \centering
  \includegraphics[width=\linewidth]{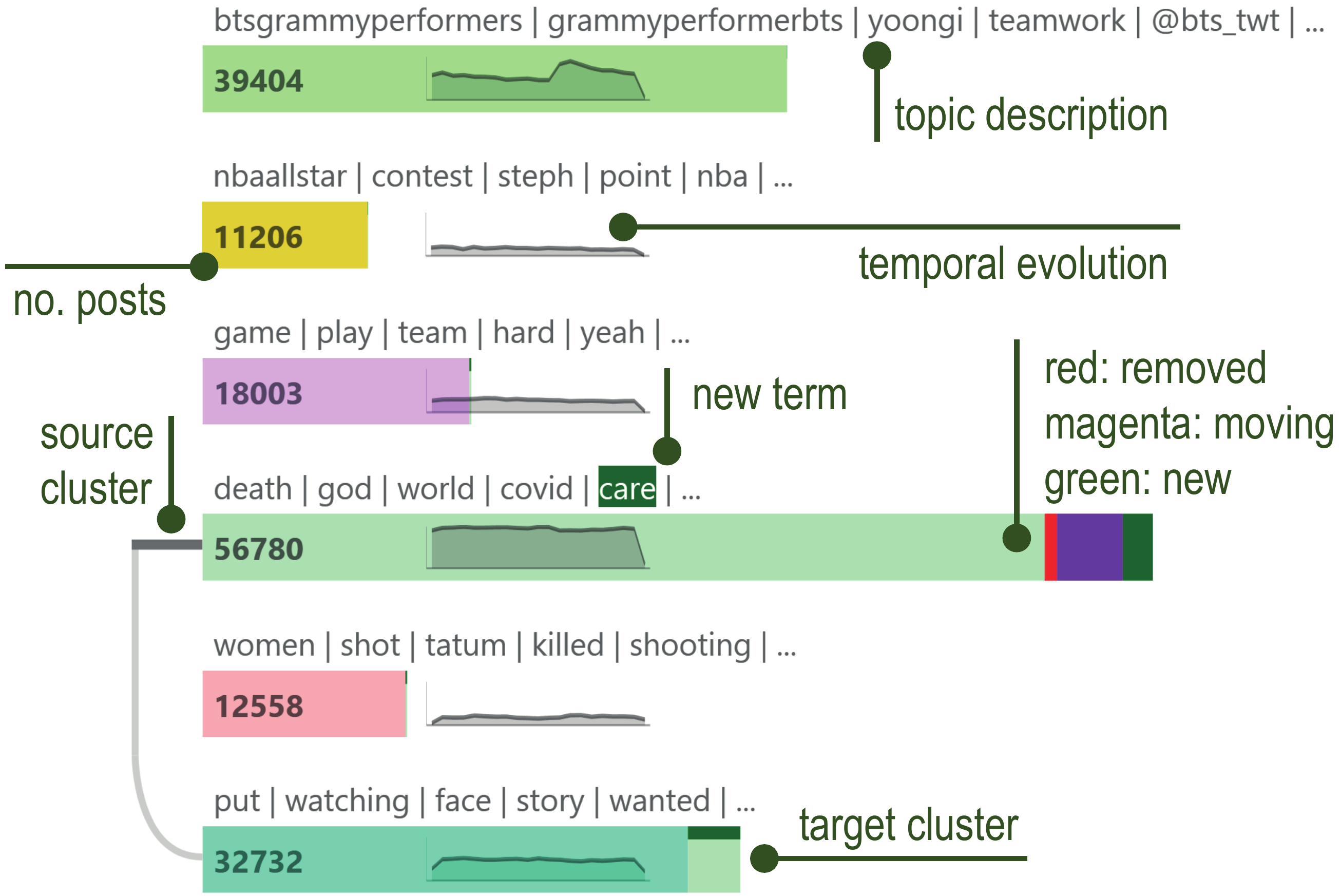}
  \caption{Overview of the topics after a new batch has been processed.
  The most defining terms of a topic serve as a description and the size of the cluster is mapped to the width of the bar below the description.
  A small line chart is embedded which shows the temporal evolution of the respective topic.
  Here, the cluster marked with the dark gray bar to the very left is currently being updated.
  The proportions of the stacked bars in red, magenta, and dark green represent the posts in the cluster that were removed, moved elsewhere, and newly added, respectively.
  Curved lines at the left side indicate to which clusters posts were moved.}
  \label{fig:topics}
\end{figure}

The resulting topics from our coarse-grained dynamic clustering process provide analysts with an interactive overview of the various themes people currently post about.
Similar to the concept of \emph{small multiples}, we plot compact summaries of the topics in a list view for an easy comparison.
Figure~\ref{fig:topics} shows an example.

One advantage of the BoW model is that the cluster centroids are interpretable.
If we sort the key-value pairs of a centroid vector in descending order of the value, we retrieve a list of the most defining terms of the respective cluster.
We take up to five of these terms to generate a descriptive but short summary of the topic's main content.
Each topic gets a distinct color and we ensure that all cluster colors have roughly the same perceived brightness, for several reasons.
This strategy mitigates perceived differences of the clusters due to dominant hues.
In addition, it makes sure that the overlays in dark gray are clearly visible.
Finally, we have a set of special colors that we use across all clusters to indicate what has changed (e.g., dark green for new posts).
These colors are darker to set them apart from the cluster colors. 
The size of each cluster is mapped to the width of its bar.
We overlay the number of posts and a small line chart onto the bar.
The line chart visualizes the temporal evolution of the cluster in the current sliding window, i.e., the number of published posts in the cluster over time.

On each update, we determine what has changed compared to the previous clustering, for instance, which posts have moved from one to another cluster.
However, revealing all changes at once might lead to a sensory overload.
Thus, we visualize the changes cluster-by-cluster from top to bottom.
A short thick line in dark gray to the left of the bar marks the current \emph{source cluster} of the update.
For instance, in Figure~\ref{fig:topics} the topic with \textit{death} as the most defining term is currently being updated.
New terms in the respective topic description are highlighted in dark green.
We further replace the bar of the current source clusters with a stacked bar to indicate the proportion of posts in the cluster that were removed from the sliding window in red, posts that have been moved to other \emph{target topics} in magenta, new posts in dark green, and the remaining posts in the original color of the cluster.
For each target topic, we append a bar that represents the proportion of posts which have been moved from the source to the respective cluster.
This bar has the same color as the source cluster, but with a dark green line at the top.
We also visualize the flow to the prevailing target topics with curves on the left side of the list.
Both the thickness \emph{and} the gray level of a curve are proportional to the square root of the number of moved posts the curve should represent.
In theory, there can be as many curves as there are topics (minus one), so we have to limit the maximum thickness.
As a result, depending on the visual encoding we would either have very thin or very light curves at times, so we use both visual variables to encode a wider range of values.

We vary the duration of each visual update depending on the complexity.
The more affected target clusters and the more appearing terms, the longer we wait before we proceed to the next cluster because users might need more time to grasp all changes.
Analysts can adjust the average speed to their needs with a toggle button at the top of the window. 
We opted for non-animated transitions to leverage visual preattentive processing so that users can immediately notice outliers and compare changes across steps more accurately.

After each update, we save the state of the topical overview in the history.
Users can choose with a slider at the bottom left of the window whether they want to peek at a previous version of the topical overview (Figure~\ref{fig:teaser}~F).
For instance, this is handy when they cannot monitor changes continuously.

Analysts can enter a search query above the list of topics (Figure~\ref{fig:teaser}~G).
Then, a new clustering session starts in which only the posts that match the query are processed.
Similarly, analysts can select one or several topics as a filter.
Both types of filters can be chained to increase the resolution down to a handful of posts.
However, only the clustering processes from the current layer are actively running.
For instance, if an analyst dove into a topic, we create a filter based on the current set of centroids at the parent layer and we use that filter for the new session, but the clustering processes of the parent level will then pause and only continue their work if the analyst goes back to the parent session.

\subsection{Frequent Phrases View}
\label{sec:frequentPhrases}

\begin{figure}
  \centering
  \includegraphics[width=\linewidth]{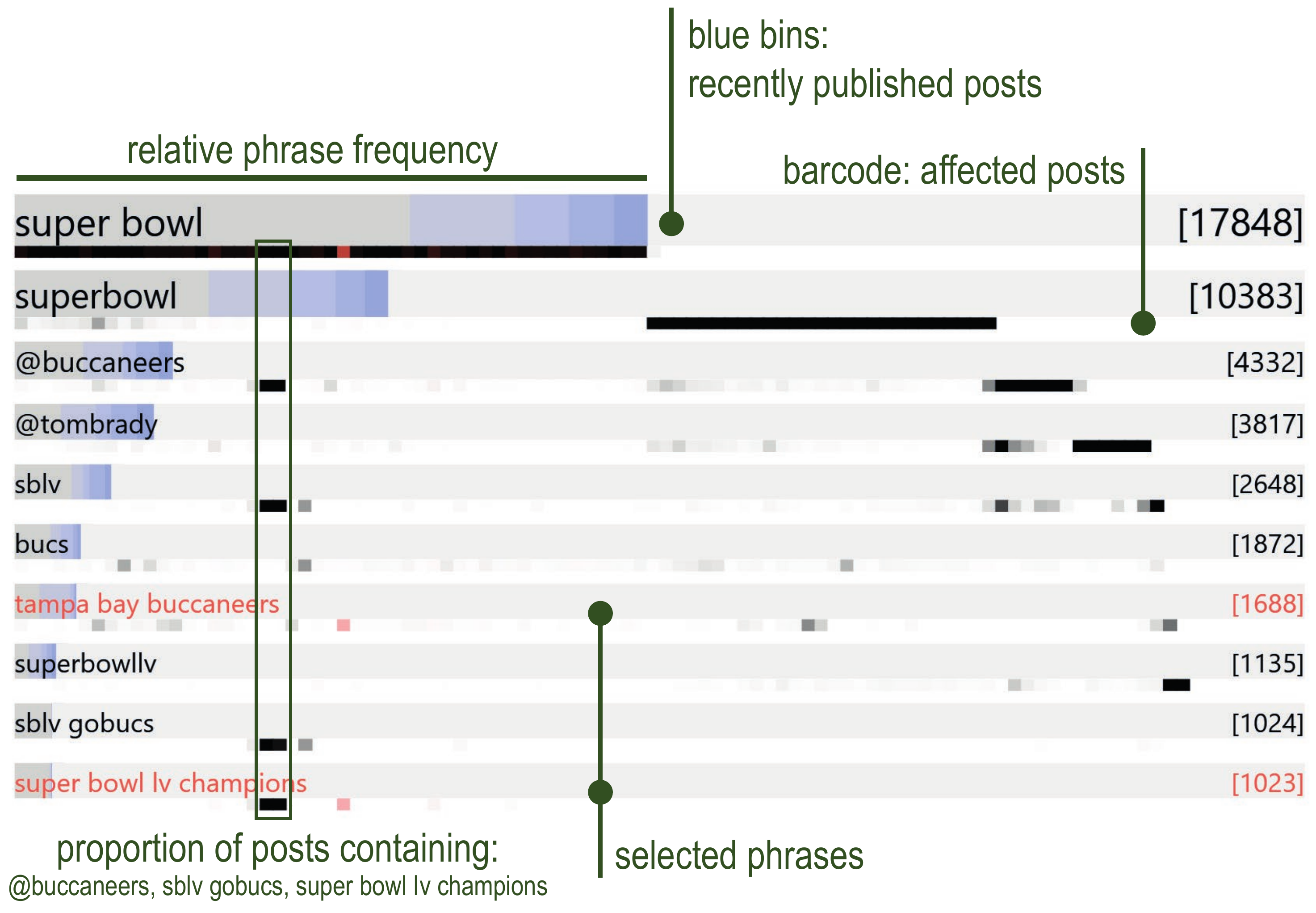}
  \caption{Visualization of the most important frequent phrases in a selection of topics.
  The bar composed of dark gray and blue-ish bins represents the number of posts containing the respective phrase.
  The blue bins depict the proportion of \emph{new} posts to indicate trends.
Each post is mapped to a horizontal position in the barcode-like visualizations below each phrase.
Dark ticks indicate the posts containing the respective phrase.
Analysts can select phrases to highlight the overlap in orange.}
  \label{fig:frequentPhrases}
\end{figure}

The short lists of terms already hint at what each topic is about, but they only offer little context.
If analysts select one or several topics of interest, we want to visualize in greater detail which issues and themes to which extent people tweet about in this topic selection.
We continuously extract the most relevant keyphrases from all tweets belonging to the selection and visualize their distribution across the posts as well as their temporal evolution.
We apply ELSKE~\cite{Knittel21Elske} for extracting the keyphrases because it belongs to one of the best performing unsupervised keyword extraction algorithms, it supports multi-term keywords (including longer phrases), and it is efficient.

Figure~\ref{fig:frequentPhrases} depicts an example of the top 10 phrases in a topic related to the super bowl event, sorted by the frequency in descending order.
The number of tweets containing the respective phrase is shown at the very right of each row and the font size of the phrase correlates with the returned importance score.
The stacked bar composed of five bins from dark gray to blue represents the proportion of tweets containing the phrase compared to all tweets in the selection.
Each bin corresponds to one fifth of the sliding window time range and depicts the proportion of tweets that were originally published within that time frame (i.e., the effective date of any retweet is the publishing date of the original tweet).
For instance, in a sliding window of 20 minutes, the blue bin at the right represents how many tweets have just been published within the last four minutes, and the dark gray bin to the left how many are older than 16 minutes.
In the example, the bars of the two phrases at the bottom are largely gray whereas the bar corresponding to \textit{@tombrady} is largely blue.
This means that people are actively tweeting \emph{new} posts containing \textit{@tombrady} at the moment but seldomly ones with \textit{super bowl lv champions} (except for retweets).
If the phrase has just appeared after an update, it will be highlighted in dark green for some seconds to catch the attention of the analyst.

Right underneath each phrase, there is a small barcode-like strip composed of 100 bins that visualizes the distribution of the corresponding phrase.
Let $n$ be the total number of posts from which the phrases were extracted.
Then, we assign each post a unique integer in the range $[1,n]$ and rescale these to real numbers in the range $[0,100)$.
The first bin then corresponds to all posts with a value in $[0,1)$, the second bin to all in $[1,2)$, and so on.
The shade of each bin from white to black represents the proportion of posts in that bin containing the respective phrase.
This helps analysts to conclude which phrases co-occur together.
For instance, most of the tweets in Figure~\ref{fig:frequentPhrases} contain \textit{super bowl} or \textit{superbowl}, but rarely both because there are only a few darker areas at the same horizontal position.
In other words, the intersection of the first two strips would be mostly white.
We greedily optimize the mapping to increase the length and number of contiguous blocks in black.
Starting with the most frequent phrase, we assign consecutive numbers to all unassigned tweets containing a certain phrase.
Let us assume we have three phrases $p_1,p_2,p_3$ (in descending order of the frequency) and $100$ tweets in total.
$40$ tweets (A) would contain the first phrase, $20$ (B) $p_2$ but not $p_1$, and $10$ (C) $p_3$ but neither $p_2$ nor $p_1$.
Then, we would assign the $40$ tweets (A) unique numbers from $0$ to $39$, tweets in B from $40$ to $59$, tweets in C from $60$ to $69$, and the remaining $30$ tweets would be assigned to slots $70$ to $99$.
Hence, tweets containing $p_2$ would never be assigned a number higher than $59$, but they could get a number lower than $40$ if they also contain $p_1$.

Analysts can click on phrases to select them.
The tweets containing all selected phrases will be highlighted in orange, and this filter then also applies to the stream of representative posts that we discuss next.

\subsection{Stream of Representative Posts}
\label{sec:representativePosts}

\begin{figure}
  \centering
  \includegraphics[width=\linewidth]{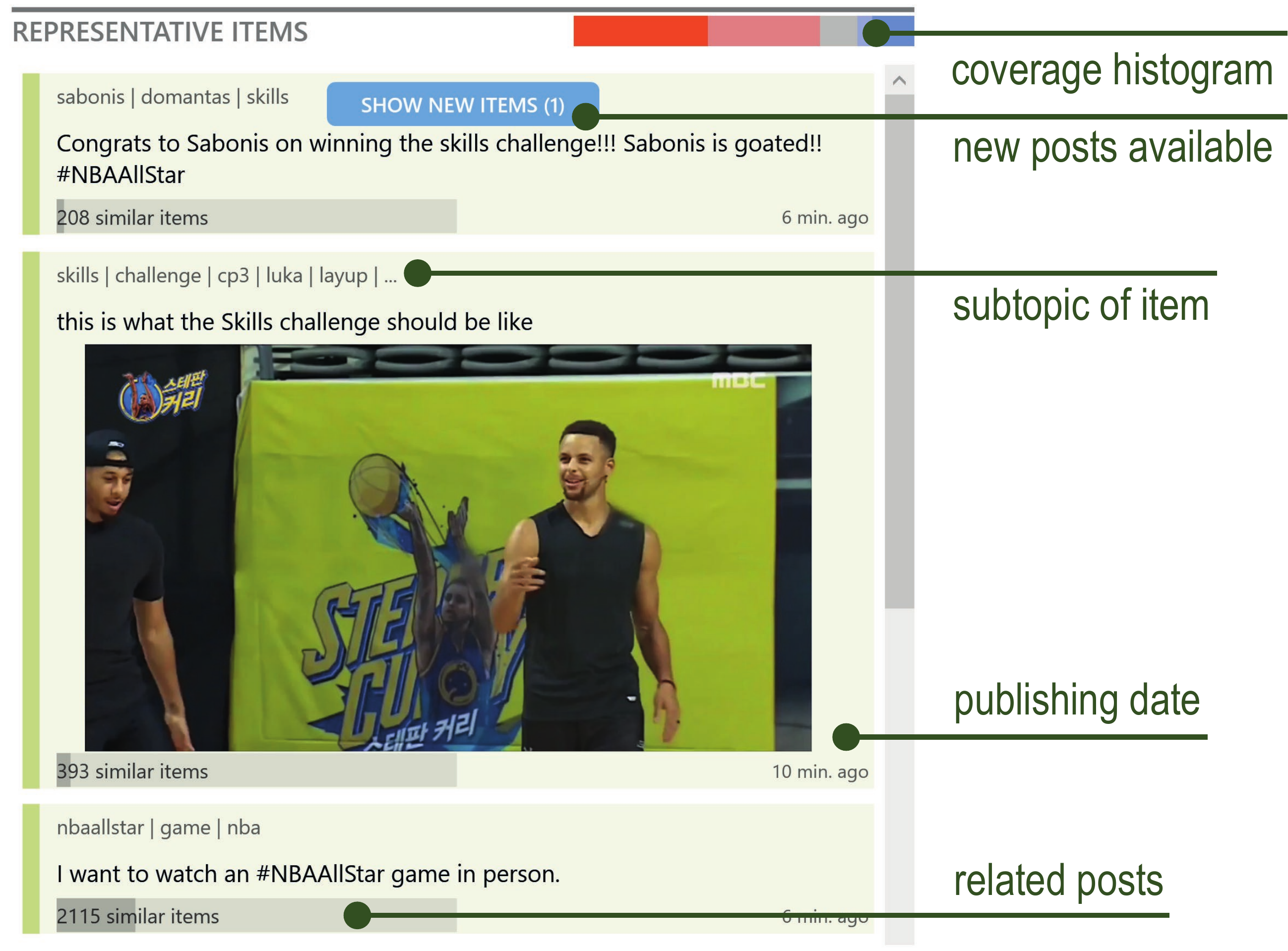}
  \caption{Stream of representative posts related to a selection of topics.
  The color of the associated topic is mapped to the background of each tweet.
  The list of terms at the top of each item describes the subtopic of that item.
  Analysts can click on the bar chart that shows the proportion of similar posts to retrieve a list of such similar posts in a new column.
  The coverage histogram at the top-right corner composed of five stacked bars indicates how well the items in the list cover all posts within the topic selection (red: proportion of similar tweets, blue: dissimilar tweets).}
  \label{fig:representativeItems}
\end{figure}

We use the fine-grained clustering process to extract representative posts for each subtopic and map them to their corresponding topic, as discussed in Section~\ref{sec:architecture}.
Inspired by the concept of \textit{learning by example}, these posts should convey the variety of points that are currently being discussed in a topic of interest.
As individual posts, they are richer in context, but they should also cover different aspects because they originated from different subtopics.
It should be noted that the number of new representative posts per update is bound by the total number of subtopics by design.
Hence, semantically zooming into topics does not increase or decrease the average number of posts in the stream, it only leads to a better coverage of the more specified topics.

Figure~\ref{fig:representativeItems} shows an exemplary stream of posts related to a selected topic.
The term list at the top of the post describes the associated subtopic.
For instance, the first tweet in the example that congratulates the basketball player for winning the skills challenge belongs to the subtopic that has \textit{sabonis}, \textit{domantas}, and \textit{skills} as the most defining terms.
The color of each tweet relates to the topic it belongs to, in case analysts select more than one topic.
At the bottom, a small bar chart depicts the proportion of tweets in the selection that are similar to the representative post (based on the cosine similarity between the document vectors).
A click on it will open a separate column with a list of related tweets, sorted by their similarity to the reference item.

If new posts have been extracted after an update, a blue button to insert these posts appears as a notification to the user.
At the top-right of the view, next to the header, stacked bars from red to blue visualize how well the extracted representations cover all posts in the selected topics (coverage histogram).
For each post in the topic selection, we calculate the cosine similarity to the representative post of the subtopic they are in.
The width of a bar then corresponds to the number of posts that have a cosine similarity in a certain range. Red represents posts with a high similarity, whereas the blue bar the ones with a low similarity, which are thus hardly covered by the stream of representative items.
If the stacked bars are mainly blue-ish, this indicates that the topic selection relates to a very diverse set of themes which are not adequately represented by the subtopics.
Analysts should then consider to increase the resolution of the analysis with a new filtered session layer.
The stacked bars in Figure~\ref{fig:representativeItems} are largely red, though, so here, the posts seem to cover most of the current content in this topic.

\section{Evaluation}

\begin{table}[tb]
\label{tab:clusterBenchmarks}
\caption{Evaluation of different clustering techniques on the \textit{20 Newsgroups} data set comprising 19,994 posts in 20 newsgroups. The normalized mutual information (NMI, higher is better) judges the quality of the resulting clustering according to the class labels. \textit{(skl)} denotes \textit{scikit-learn}-based implementations. The results show that our spherical k-Means++ implementation (in bold) is not only fast but also leads to the best clustering result compared to LDA, NMF, and k-Means++.}
\centering
\begin{tabu}{@{}lrr@{}}
\toprule
                             & \multicolumn{1}{c}{\textbf{NMI}} & \multicolumn{1}{c}{\textbf{Duration}} \\ \midrule
LDA \textit{(skl)}           & 0.19 {[}0.18 - 0.20{]}          & 314.5s {[}313.1s - 315.5s{]}             \\ \hline
NMF \textit{(skl)}           & 0.48 {[}0.48 - 0.49{]}          &  22.3s {[}22.2s -22.4s{]}             \\ \hline
k-Means++ \textit{(skl)}     & 0.46 {[}0.45 - 0.47{]}          & 20.0s {[}17.4s - 20.3s{]}             \\ \hline
k-Means++                    & 0.45 {[}0.44 - 0.46{]}          &  \textbf{4.1s} {[}3.2s - 4.8s{]}             \\ \hline
\textbf{spherical k-Means++} & \textbf{0.56} {[}0.56 - 0.57{]}          &  4.2s {[}3.7s - 4.6s{]}      
         
\end{tabu}
\end{table}

\begin{table*}[]
\label{tab:streamingBenchmarks}

\caption{Evaluation of our sliding window-based dynamic clustering approach on the \textit{20 Newsgroups} data set that was processed as a batched stream. Each batch contains about 2k documents. The resulting NMI refers to the final batch (higher is better, interquartile range in brackets). The clustering coherence score captures the average similarity of centroids between subsequent runs. Our approach (in bold) leads to better and more coherent clustering results compared to processing the bins on their own.}
\centering
\begin{tabu}{@{}lrrr@{}}
\toprule
                                         & \multicolumn{1}{c}{\textbf{NMI}} & \multicolumn{1}{c}{\textbf{Coherence}} & \multicolumn{1}{c}{\textbf{Duration}} \\ \midrule \vspace{6pt}
Baseline (sKMeans++, distinct bins)                 & 0.50 {[}0.41 - 0.50{]}           & 0.37 {[}0.37 - 0.38{]}                 & 0.13s {[}0.12s - 0.13s{]}             \\ 
\textit{10 clusters max.}                & \multicolumn{1}{l}{}             & \multicolumn{1}{l}{}                   & \multicolumn{1}{l}{}                  \\
\textbf{Dyn. sKMeans++, 75\% overlap} & 0.62 {[}0.61 - 0.63{]}           & \textbf{0.62} {[}0.60 - 0.62{]}                 & \textbf{0.11s} {[}0.11s - 0.12s{]}             \\ \hline
\textbf{Dyn. sKMeans++, 50\% overlap} & \textbf{0.63} {[}0.61 - 0.63{]}           & 0.61 {[}0.61 - 0.62{]}                 & 0.14s {[}0.14s - 0.15s{]}             \\
\\ \vspace{3pt}
\textit{20 clusters max.}                & \multicolumn{1}{l}{}             & \multicolumn{1}{l}{}                   & \multicolumn{1}{l}{}                  \\
\textbf{Dyn. sKMeans++, 75\% overlap} & 0.57 {[}0.57 - 0.58{]}           & 0.53 {[}0.50 - 0.53{]}                 & 0.18s {[}0.17s - 0.18s{]}             \\ \hline
\textbf{Dyn. sKMeans++, 50\% overlap} & 0.59 {[}0.59 - 0.59{]}           & 0.54 {[}0.53 - 0.54{]}                 & 0.22s {[}0.22s - 0.23s{]}            
\end{tabu}
\end{table*}

We compared the clustering quality and efficiency of popular clustering algorithms (Section~\ref{sec:clusteringBenchmarks}) and evaluated our dynamic clustering algorithm on the \textit{20 Newsgroups} data set (Section~\ref{sec:streamingBenchmarks}).
In Section~\ref{sec:useCases}, we discuss two use cases to show the utility of our approach.

\subsection{Clustering Benchmarks}
\label{sec:clusteringBenchmarks}

We adapted the spherical k-Means algorithm to our dynamic clustering approach because it has two important benefits.
It belongs to one of the fastest clustering algorithms and previous work indicates that it performs well on document collections~\cite{lelu:hal-03053176}.
To corroborate these assumptions, we evaluated several popular document clustering methods with the well-known \textit{20 Newsgroups} data set\footnote{http://qwone.com/~jason/20Newsgroups/}.
It contains nearly 20,000 posts spread across 20 different newsgroups, and the corresponding newsgroup of each post serves as a class label.
It is difficult to obtain ground truth labels since the grouping may also depend on individual preferences and the task at hand.
Nevertheless, the class label allows us to judge how well the clustering results match the crowd-sourced association of the documents with one of the twenty categories.

\subsubsection{Test Setup}

We converted each document (that is, the \textit{Subject} line and the actual body) into a TF-IDF-weighted BoW representation.
The inverse document frequency is based on the complete data set.
We ignored stop words and normalized each vector to have unit length, but apart from stop words, we did not truncate the vocabulary.
Three posts were excluded because they only contain stop words, so the final input for the clustering comprises 19,994 non-zero vectors in total.
We ran each algorithm five times to reduce the impact of outliers.

In addition to testing our own (spherical) k-Means++ implementations, we used the popular \texttt{scikit-learn}\footnote{https://scikit-learn.org} implementations (SKL) of Latent Dirichlet Allocation (LDA), Non-Negative Matrix Factorization (NMF) and k-Means++.
The number of clusters was set to $20$ and the remaining (hyper-)parameters were kept to their defaults, with only one exception.
With default settings, the \texttt{scikit-learn} implementation of k-Means++ runs the algorithm ten times and returns the clustering with the lowest distortion.
For a fair comparison with the remaining methods, we set \texttt{n\_init} to $1$ so that the algorithm runs only once upon each invocation.
We tested the k-Means++ algorithm twice (SKL and ours) to make sure that our implementations are compatible and comparable with the \texttt{scikit-learn} setup.
All experiments were run on the same device equipped with a six-core desktop CPU, and all algorithms used the exact same (sparse) input matrix.

\subsubsection{Results}

Table~\ref{tab:clusterBenchmarks} lists the results.
We report the median duration and normalized mutual information (NMI) of each run, with the respective interquartile ranges in brackets.
The NMI is an information-theoretic-based external criterion to judge how well a clustering matches the class labels.
It ranges between $0$ (no correlation) and $1$ (perfect correlation).
One advantage of this score is that it also works if the number of classes or clusters differs between the two sets.

The results show that the spherical k-Means++ algorithm is not only fast, it also clusters the data set best according to the NMI score. 
NMF and (euclidean) k-Means++ perform similarly and slightly worse than the spherical k-Means++ version that uses the cosine distance.
LDA, however, takes several minutes to complete and leads to a lower-than-average result, which is in line with previously reported results~\cite{lelu:hal-03053176}.
Hence, our experiments corroborate the finding that the k-Means algorithm using the cosine distance clusters document collections reasonably well.
It should be noted that our implementation is even more efficient if applied to tweets.
Clustering 100,000 tweets with $k = 100$ takes around two seconds on the same system.
As described in Section~\ref{sec:clustering}, we exploit the sparsity of input vectors, and tweets are only composed of about $20$ tokens on average.

\subsection{Dynamic Clustering Benchmarks}
\label{sec:streamingBenchmarks}

We ordered the posts in the \textit{20 Newsgroups} data set by their publishing date and simulated a streaming environment to investigate how well our dynamic clustering approach (\textit{dyn. sKMeans++}) performs.

\subsubsection{Test Setup}

The data set was processed in batches of approximately 2,000 posts (10\% of the data set size) with two different strategies: our dynamic clustering approach and a baseline for comparison.
We calculated the normalized mutual information (NMI) on the final batch with the corresponding labels to judge the quality of each clustering.
For evaluating the \textit{coherence} between two clusterings and their corresponding centroid sets $C_1$ and $C_2$, we calculated the average cosine similarity of the centroids in $C_1$ with their corresponding closest match in $C_2$ and vice versa.
We ran each strategy and configuration five times.
The setup of each strategy was as follows:

\textbf{Baseline (distinct bins):} The data set was split into 10 distinct batches and our spherical k-Means++ algorithm was run on each batch separately, with $k$ set to the number of ground-truth classes in this batch.
The reported \textit{Coherence} score is the average coherence between all pairs of subsequent batches.

\textbf{Our approach (dyn. sKMeans++):} We applied our dynamic spherical k-Means++ algorithm to the data set with a sliding window size equal to the batch size of the baseline scenario.
As the window slides forward, new posts are added and old ones removed.
We tested two strides, one that leads to an overlap of $75\%$ between subsequent windows and one that leads to $50\%$ overlap.
In contrast to the baseline scenario, the number of ground-truth classes was not fed to the algorithm.
Rather, we evaluated two configurations with a maximum of 10 and 20 possible clusters, respectively.
Here, the \textit{Coherence} score is the average coherence between all pairs of subsequent \emph{distinct} batches to allow a fair comparison with the baseline scores.
For instance, with a stride of one-fourth of the window size ($75\%$ overlap), we calculate the coherence between batches $1$ and $5$, $5$ and $9$, and so on.

\subsubsection{Results}

Table~\ref{tab:streamingBenchmarks} lists the results.
Our approach leads to better and more coherent clustering results in all configurations compared to the baseline, and the overall duration of each step does not increase, despite additional optimization runs for the dynamic version.
Hence, taking the previously calculated centroids into account in the initialization step of the algorithm has several benefits.
It leads to more coherent clustering results between subsequent updates, it leads to faster convergence within a single optimization run, and it also leads to better clustering results on the \textit{20 Newsgroups} data set.
The higher NMI scores may seem surprising at first, but one reason for this finding is that the initialization strategy accumulates to some extent knowledge of previous batches, which improves the generalizability on new data.

\subsection{Use Cases}
\label{sec:useCases}

\begin{figure*}
  \centering
  \includegraphics[width=\linewidth]{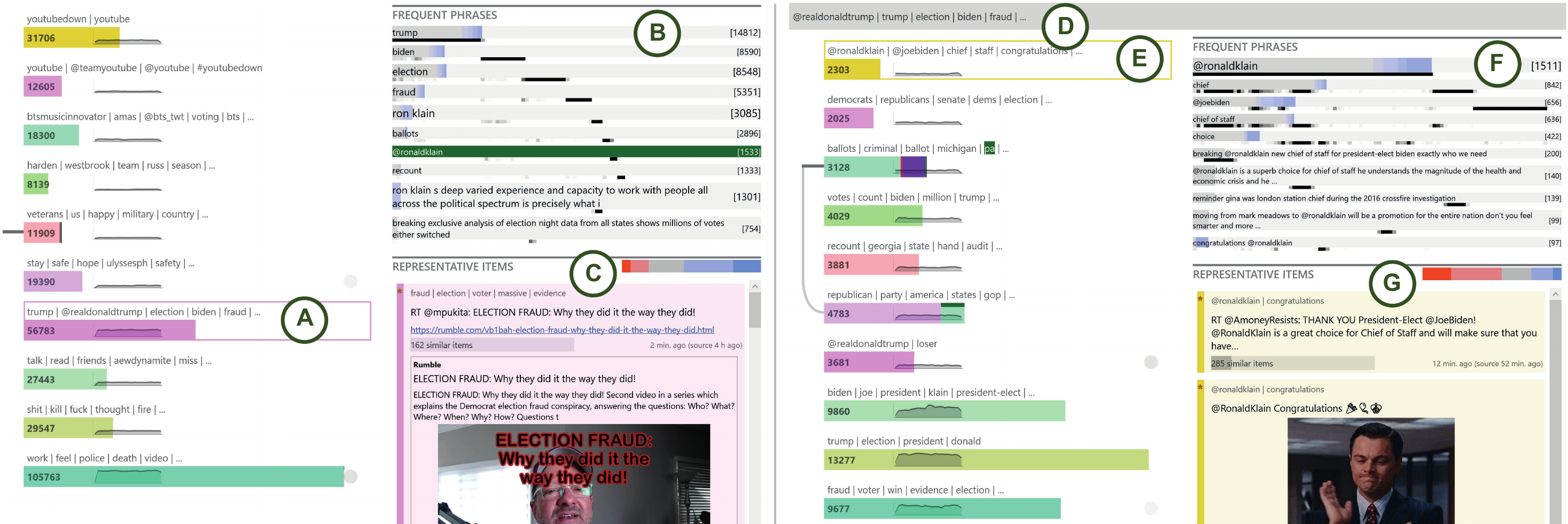}
  \caption{\textbf{Left:} top-level view with one selected topic (A) and the corresponding frequent phrases (B) and representative posts (C).
  \textbf{Right:} sub-level view using filtered session (D) based on top-level topic (A), with selected topic (E) and its frequent phrases (F) and representative posts (G).}
  
  \label{fig:diveInUseCase}
\end{figure*}

\subsubsection{NBA, BTS, and Oprah}

In this use case, we look at tweets that were streamed on Sunday evening (US ET), March 7th, 2021, at a rate of about 250 new posts per second.
The size of the sliding window is 20 minutes.
The analyst first scans all topics and notices that people currently seem to tweet about the Grammys, the NBA All Stars Event, Covid, the AEW Revolution Wrestling Event, and other more general topics.
One bigger topic that includes the hashtag \textit{btsgrammyperformers} catches the interest of the analyst who clicks on it to find out more.
From the visualized distribution of the frequent phrases they conclude that this is a homogeneous cluster in which people mostly tweet about the fact that the boygroup BTS was announced to perform at the Grammys the week after, including celebrations of Min Yoon-gi who is a member of BTS.

The analyst notices that the line chart of the \textit{nbaallstar, steph, curry} cluster indicates a strong upward trend.
After switching to that topic, the list of representative items reveals that thousands of the received tweets cheer on the basketball player Steph Curry.
After a while, the blue badge appears, notifying the analyst that new posts are available.
They click on it to insert the recently extracted tweets into the list.
Most of the new tweets congratulate him (\textit{Steph Curry greatest shooter ever!}), again with thousands of similar tweets, indicating that the basketball player is performing very well in the currently running game. 

Later, a new cluster appears about Meghan Markle at Oprah (\textit{harryandmeghanonoprah}).
Initially, people talk about whether and where to watch the interview.
Shortly afterward, the line chart of the topic goes strongly up (Figure~\ref{fig:teaser}).
From the visualization of the frequent phrases (C), the analyst concludes that the social media community has just started to increasingly post about Meghan giving an interview to Oprah because the phrases have largely blue-ish backgrounds.
One representative post (E') talks about an utterance in the interview that it was apparently Kate who made Meghan cry.
The analyst clicks on the bar in the lower-left to retrieve related tweets (Figure~\ref{fig:teaser} right column).
From the tweets, it becomes clear that most users make fun of the fact that Meghan goes after Kate, even though some seem to be very upset (\textit{It's Kate I'm sure}).
Later on, some tweets have a more serious tone after Meghan talks about racism (\textit{How dark would Archie's skin be?!!}).

This use case shows that our system not only makes analysts aware of major topics that are discussed on social media, but it also enables the specific monitoring of ongoing events, even if the frequency of posts suddenly increases.
It also shows that our clustering-based approach helps to differentiate between several major events happening at the same time.
In addition, the topic descriptions clearly indicate that considering novel terms in our pipeline (e.g., \textit{btsgrammyperformers}) is beneficial for the clustering of tweets.

\subsubsection{YouTube Outage and Dive Into Politics}

This use case deals with streamed tweets from November 11th, 2020.
The left part of Figure~\ref{fig:diveInUseCase} depicts the top-level view.
The topic description of one of the first topics that catches the interest of the analyst contains only two major terms: \textit{youtubedown} and \textit{youtube}.
The analyst hypothesizes that YouTube is experiencing some kind of outage and selects the topic to investigate their hypothesis.
The visualization of the frequent phrases confirms that nearly all of the more than 30,000 posts in this topic indeed contain \textit{youtubedown}.
A large proportion of the phrase background is composed of blue-ish bars, which indicates an ongoing issue since many people are actively posting new tweets and not just retweets.
The analyst notices that there is another similar topic about \textit{youtube} and adds it to the current selection.
The combined stream still contains only a handful of posts, but the large red bar in the coverage histogram next to the header indicates that these posts cover most of the published tweets well.
Thus, they conclude that most posts are slight variations of the utterance that YouTube is down.


Then, the analyst looks at the topic about Biden, Trump, and the presidential election (Figure~\ref{fig:diveInUseCase} A).
However, the coverage histogram (C) reveals that the topic is relatively diverse because the stream does not adequately represent a significant proportion of posts.
They click on the Dive-In button which is situated at the upper-left of the window to start a new filtered session.
The Topical Overview now relates to topics derived from clusters of the parent topic, as shown in the right part of Figure~\ref{fig:diveInUseCase}.
The analyst first focuses on the \textit{@ronaldklain} topic (E) and learns that Joe Biden has picked Ronald Klain as Chief of Staff.
There are many congratulations (G), but also critical remarks about his alleged handling of the swine flu.
It should be noted that the visual patterns between the frequent phrase visualization of the top-level topic (B) and the sub-level topic (F) clearly differ.
Whereas the step-like pattern in (B) reveals that the different keywords seldomly appear together, the pattern in (F) shows significant overlaps of keyphrases and, thus, points toward a more homogeneous topic.
Furthermore, the proportion of the two red-like bars in the coverage histogram (G) is much larger compared to (C).
The analyst now skims through the other topics in which people talk about the ballots (\textit{many faked ballots in detroit}, \textit{Where ARE those ballots?}), potential voter fraud, the recount by hand in Georgia that the secretary of state seems to have just announced, but also Democrats celebrating the win.

This use case demonstrates that our visualization of frequent phrases and the coverage histogram enable analysts to assess the composition of topics and the diversity of relevant posts.
It also highlights the utility of our layered approach for semantically zooming into topics of interest.

\section{Discussion}

Compared to previous work, our approach offers several advantages.
First, it can handle a large number of posts even on a budget PC.
We successfully tested our system with a sliding window size of more than two million posts.
Second, it visually structures the data without relying on additional metadata or any kind of pre-filtering.
Third, for a selection of topics, analysts can retrieve a real-time stream of representative posts irrespectively of the actual frequency of published posts.
With the visualization of the frequent phrases, analysts can assess what the topic is about, how homogeneous it is, and whether it is currently going viral.
Fourth, our approach is agnostic as to which social media platform is analyzed because it only processes the content of the posts.
Finally, we tested our system with different languages, including Spanish and German, to verify that it generalizes to non-English languages.

This work also has limitations.
Our pipeline is purely content-based and enables top-down analyses, so analysts may miss smaller developments with just a handful of associated posts and retweets.
People also increasingly post images or videos with only a short description, which reduces the effectiveness of our clustering algorithm because we currently ignore media content for efficiency reasons.
Another important aspect is the homogeneity of the posts.
In the case of major events, a large proportion of published posts belong to these events. 
The everyday content is much more diverse, though.
Hence, it may be difficult to group such posts in a meaningful way on a coarse level.
Analysts would then either need to increase the maximum number of topics or dive into one of the less focused topics to find interesting themes.
Users might also want to include tweets from different languages regarding an ongoing event.
We could extend our system such that it determines the inverse document frequency from different reference corpora, based on the language of each tweet.

\section{Conclusion and Future Work}

We presented an interactive system for the visual analysis of streaming social media posts.
Compared to previous work, our system enables a fast and comprehensive analysis of larger data sets in real-time and, thus, contributes to making the visual analysis of streaming documents more scalable.
The use cases indicate that our system not only supports analysts in getting an overview of what is currently happening on social media platforms but also in monitoring specific topics at different resolutions.
The benchmarks show that our clustering algorithm performs well on documents, is efficient, and optimizes the coherence of clusters between updates to preserve the mental map of analysts.

We aim to improve the analysis workflow with pinned filters in the near future.
If analysts have zoomed into a topic of interest, they could pin this configuration of chained filters as a new classifier so that they can go back to a higher level while still being able to monitor that specific topic.
Furthermore, we would like to investigate how we can integrate optionally available metadata, as well as shared links and images, into the analysis process.

\acknowledgments{
This research was supported by the German Science Foundation (DFG) as part of the project VAOST (392087235) and as part of the Priority Program VA4VGI (SPP 1894).
It was also partially funded by the joint Sino-German program of the NSFC (61761136020).}

\bibliographystyle{abbrv-doi}

\bibliography{streaming-cluster-vis-paper}
\end{document}